\documentclass[aps,pra,showpacs,twocolumn,amsmath,amssymb,superscriptaddress, footinbib]{revtex4}

\usepackage[english]{babel}

\usepackage{latexsym}
\usepackage{graphicx}
\usepackage{subfigure}
\usepackage{epsfig}
\usepackage{amsfonts}
\usepackage{amssymb}
\usepackage{amsmath}
\usepackage{bm, bbm}

\usepackage{lipsum} 
\usepackage[draft]{todonotes}   

\begin{document}

\title{Disordered cold atoms in different symmetry classes} 

\author{Fernanda Pinheiro}
\affiliation{Department of Physics,
Stockholm University, AlbaNova University Center, 106 91 Stockholm,
Sweden}
\affiliation{Institut f\"ur Theoretische Physik, Universit\"at zu K\"oln,
De-50937 K\"oln, Germany}
\affiliation{NORDITA, KTH Royal Institute of Technology and Stockholm University, Se-106 91 Stockholm, Sweden}
\author{Jonas Larson}
\affiliation{Department of Physics,
Stockholm University, AlbaNova University Center, 106 91 Stockholm,
Sweden}
\affiliation{Institut f\"ur Theoretische Physik, Universit\"at zu K\"oln,
De-50937 K\"oln, Germany}
\date{\today}

\begin{abstract}
We consider an experimentally realizable model of non-interacting but randomly coupled atoms in a two-dimensional optical lattice. By choosing appropriate real or complex-valued random fields and species-dependent energy offsets, this system can be used to analyze effects of disorder in four different classes: The chiral BDI and AIII, and the A and AI symmetry classes. These chiral classes are known to support a metallic phase at zero energy, which here, due to the inevitable finite size of the system, should also persist in a neighborhood of non-zero energies. As we discuss, this is of particular interest for experiments involving quenches. Away from the centre of the spectrum, we find that excitations appear as domain walls in the cases with time-reversal symmetry, or as vortices in the cases where time-reversal symmetry is absent. Therefore, a quench in a system with uniform density would lead to the formation of either vortices or domain walls depending on the symmetry class. For the non-chiral models in the A and AI classes, a population imbalance between the two atomic species naturally occurs. In these cases, one of the two species is seen to favour a more uniform density. We also study the onset of localization as the disorder strength is increased for the different classes, and by deriving an effective model for the non-chiral cases we show how their eigenstates remain extended for larger values of the coupling with the disorder, if compared to the non-chiral ones.
\end{abstract}

\maketitle

\section{Introduction} 
Symmetries play a fundamental role in physics, ranging from predicting planetary motion to understanding interaction among elementary particles. Indeed, properties of quantum mechanical systems in equilibrium are often universal and can be characterized according to symmetries~\cite{sym,haake}. Recently, a more in-depth understanding for the aspects of various lattice models have been advanced thanks to the development of topological insulators~\cite{top}. Not surprisingly, the characteristics of such lattice models rely on the underlying symmetries, i.e. time-reversal, particle-hole, and chirality. Structuring the models accordingly yields  a ``periodic table'' with ten different symmetry classes~\cite{altland}. 

Together with the dimensionality, these symmetry classes tell much about the properties of a system. Thus, a question that naturally arises is: Can we identify physical systems that allow for control of symmetry classes? Then, by turning a knob the experimentalist could in principle qualitatively alter the system's properties. In this respect, systems of cold atoms loaded into optical lattices are extremely attractive~\cite{rev}. These are well isolated from any environments, clean from impurities, and preparation and detection are relatively easy. The lattice, geometry and dimensionality, can be monitored with great flexibility, and by using multiple atomic electronic states one can realize multi-species models~\cite{rev}.  Furthermore,
cold atomic systems also allow for a systematic study of effects deriving from disorder
by handling the type and strength of the disorder with additional laser  fields~\cite{atomloc}.

Back in 1958,  Anderson predicted that disorder can fully inhibit transport due to disorder induced coherent scattering resulting in destructive quantum interferences~\cite{anderson}. This absence of conductance, following the appearance of spatially localized eigenstates in the system, is the phenomenon of {\it Anderson localization} or strong localization. In one dimension (1D), arbitrarily weak disorder localizes every eigenstate in the entire spectrum, while in 3D there exists a `mobility edge' separating localized from extended (or metallic) states~\cite{anderson2}. Thus, in 3D it is possible to realize a metal-Anderson insulator transition by tuning the disorder strength. In 2D, the situation becomes more intriguing because the presence or absence of insulating and/or metallic phases strongly relies on the symmetry classes~\cite{2ddis}. The eigenstates of chiral systems in 2D, for example, may become 
delocalized at the center of the spectrum~\cite{rfm1}. But even if this phenomenon is accepted today, the existence of delocalized states in such systems was for 
long debated~\cite{rfm}. 
 
In this work we study a cold atom model which is of easy experimental implementation. We consider two non-interacting atomic species coupled by random fields and confined in an optical lattice. Physically the coupling amounts to a laser induced Raman coupling between two internal atomic Zeeman levels. By adjusting the properties of the coupling lasers, i.e. the phase and frequency, the system may fall in four different symmetry classes, where two supports a metal-Anderson insulator transition at zero energy. Properties of the eigenstates are explored numerically, and it is found that, depending on the symmetries, the excited states can host domain-walls or vortices. The vortices appear in the cases with a complex-valued Raman coupling, i.e. broken time-reversal symmetry. For the system with chiral symmetry, the model can be mapped onto a {\it random flux model}~\cite{rfm1,rfm} where every lattice plaquette is subject to a (random) synthetic magnetic field. When the chiral symmetry is broken, the two species are reorganized such that one of them 
carries the system's kinetic energy, being characterized, therefore, by a smooth density, while the other one shows a vivid structure that follows the random Raman coupling corresponding to lowering the potential energy. This behavior is explained in terms of an effective model valid in the limit of a large population imbalance between the two species. We also discuss the model in light of the Mermin-Wagner theorem~\cite{mw} which impose restrictions on the establishment of long range order, depending on the symmetries and the dimensions of the system.
This is in connection to the phenomenon known by the name of {\it random field induced order} (RFIO)~\cite{rfio}, in which a particular choice of disorder can be used to lower the symmetries of the clean system and consequently, to invalidate the premisses under which the Mermin-Wagner theorem can be applied, thereby stabilizing long-range order. Despite clear similarities with models studied in the past that have been shown to exhibit RFIO~\cite{armand} we do not find evidence of RFIO in the non-interacting case studied here. 
     
The paper is structured as follows. In the next section we introduce the model system and the corresponding Hamiltonian in a general form. The symmetries and the different classes are presented in Subsec.~\ref{symmetrysec}. After summarizing known results on the universal properties of the symmetry classes discussed here in Sec.~\ref{resultsec} we present the numerical results: We characterize the eigenstates in Subsec.~\ref{propsubsec}, and then the RFIO in Subsec.~\ref{rfiosubsec}. In the last section~\ref{sec:con} we give a summary of the results and briefly touch upon the question how to extend the system to other classes.
  
\section{System}\label{Systemsec}
The controllability of cold atomic systems together with the possibility to at will determine the disorder properties have made these systems a perfect testbed to study effects like Anderson localization~\cite{atomloc}. In the early experiments, 1D tubes of cold (bosonic) atoms were explored and either speckle potentials~\cite{atomdis1} or incommensurate optical lattices~\cite{atomdis2} were used to realise disorder (or quasi-disorder). In these pioneering early cold atom experiments it was possible to extract both the hindered matter wave spreading and the localization length.
Influence of interaction was later probed and a crossover from localized to delocalized states could be observed~\cite{inguscio2}. At a mean-field level, the non-linearity stemming from the atom-atom interaction effectively couples localized states leading to a delocalization~\cite{gploc}. In 3D, Mott argued for the presence of a mobility edge separating localized states at the tails of the spectrum, from extended ones at the centre~\cite{Mott}. This has recently been verified using both fermionic~\cite{fermion3d} as well as bosonic atoms~\cite{boson3d}. In the Bose-Hubbard (BH) model, a suppression of superfluidity has been seen in 3D~\cite{BHdis}. Further investigations~\cite{BHglas} suggested this as a signature of the Bose glass phase predicted years ago for this model~\cite{fisher}.
 
In 2D, more relevant for this work, the glass phase of the disordered BH model has been explored as well as interaction driven transitions from a localized phase to a superfluid phase and finally a reentrance into a Mott insulating phase~\cite{2dint}. The (repulsive) interaction counteracts the disorder induced localization, and for strong enough interaction the atoms forms a disordered Mott insulator state. For non-interacting gases, the diffusion in a disordered 2D optical lattice was measured in Ref.~\cite{2ddiff}. By applying the speckle potential with an incident angle the disordered potential shows a different correlation length in the two direction which led to anisotropic spreading. In terms of the symmetry classes, this experimental system falls within the AI class (see below) where all the states are known to be localized, and furthermore, this class does not seem to support topological states. Following this rapid progress, we continue by suggesting a experimentally simple model which goes beyond present experiments and show new characteristics.

\subsection{Model Hamiltonian}\label{modelham}
\begin{figure}[h]
\centerline{\includegraphics[width=6.7cm]{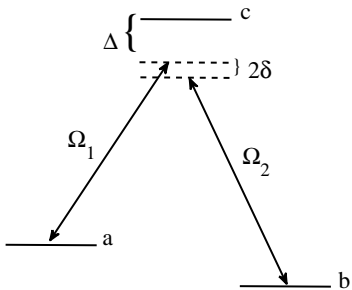}}
\caption{Schematic plot of the atom-laser $\Lambda$ coupling configuration. The states $|a\rangle$ and $|c\rangle$ are dipole coupled by a laser with (spatially dependent) amplitude $\Omega_1$, and a second laser couples the states $|b\rangle$ and $|c\rangle$ with an amplitude $\Omega_2$. The atom-laser detuning $|\Delta|\gg|\Omega_1|,\,|\Omega_2|$ such that the excited state $|c\rangle$ becomes only virtually populated. Thus, it is adiabatically eliminated resulting in an effective coupling between the states $|a\rangle$ and $|b\rangle$ with an amplitude $\Omega=\Omega_1^*\Omega_2/\Delta$. $2\delta$ is the detuning for the two-photon process.} \label{fig1}
\end{figure}

We consider non-interacting three-level atoms, with internal states labeled $a$, $b$, and $c$, confined to a 2D square optical lattice. The internal electronic states $|a\rangle$, $|b\rangle$ and $|c\rangle$ are Raman-coupled with two external lasers as explained in Fig.~\ref{fig1}. By introducing the atomic field operators $\hat\Psi_\alpha({\bf x})$ ($\alpha=a,\,b,\,c$), that obey the regular bosonic commutations
\begin{equation}
\left[\hat\Psi_\alpha({\bf x}),\hat\Psi_\beta^\dagger({\bf x}')\right]=\delta_{\alpha\beta}\delta({\bf x}-{\bf x}'),\hspace{0.5cm}\left[\hat\Psi_\alpha({\bf x}),\hat\Psi_\beta({\bf x}')\right]=0,
\end{equation}
the Hamiltonian is given, in the rotating wave approximation~\cite{com1}, by
\begin{widetext}
\begin{equation}
\begin{array}{lll}
\hat{\mathcal{H}} & = & \displaystyle{\int\,d{\bf x}\Bigg\{\sum_{\alpha=a,b,c}\hat\Psi_\alpha^\dagger({\bf x})\hat H\hat\Psi_\alpha({\bf x)}+\Delta\hat\Psi_c^\dagger({\bf x})\hat\Psi({\bf x})+\delta\hat\Psi_a^\dagger({\bf x})\hat\Psi_a({\bf x})-\delta\hat\Psi_b^\dagger({\bf x})\hat\Psi_b({\bf x})}\\ \\
& & +\left[\Omega_1({\bf x})\hat\Psi_c^\dagger({\bf x})\hat\Psi_a({\bf x})+h.c.\right]+\left[\Omega_2({\bf x})\hat\Psi_c^\dagger({\bf x})\hat\Psi_b({\bf x})+h.c.\right]
\Bigg\}.
\end{array}
\end{equation}
\end{widetext}
Here, $\hat H=-\frac{\hbar^2}{2m}\left(\frac{\partial^2}{\partial x^2}+\frac{\partial^2}{\partial y^2}\right)+V_0\left[\cos^2(kx)+\cos^2(ky)\right]$ is the 2D lattice Hamiltonian with $m$ the atomic mass, $k$ the wave vector (the same in both directions), and $V_0$ the potential amplitude (also equal for both directions). $\Delta$, $\delta$ and $\Omega_i({\bf x})$ ($i=1,\,2$) are atom-light detunings and drive amplitudes as explained in Fig.~\ref{fig1}, and in particular,
the two amplitudes $\Omega_i({\bf x})$ are in general complex (which can be controlled by the phases of the lasers). 

\subsection{Properties}\label{propsubsec}

We assume the dispersive coupling regime $|\Delta|\gg|\Omega_1({\bf x})|,\,|\Omega_2({\bf x})|,\,|\delta|$ meaning that if the internal states of the atoms are initialized in the lower states $|a\rangle$ and $|b\rangle$, the excited state $|c\rangle$ is only weakly populated and can be integrated out. In this regime, the field
\begin{equation}
\hat\Psi_c({\bf x})=-\frac{\Omega_1({\bf x})\hat\Psi_a({\bf x})+\Omega_2({\bf x})\hat\Psi_b({\bf x})}{\Delta}.
\end{equation}
is assumed to follow
the other two fields adiabatically~\cite{molmer}. Within this assumption, we derive an effective model for the remaining fields~\cite{com2}
\begin{equation}
\begin{array}{lll}
\hat{\mathcal{H}}_\mathrm{eff} & = & \displaystyle{\!\!\int d{\bf x}\Bigg\{\!\sum_{\alpha=a,b}\!\hat\Psi_\alpha^\dagger({\bf x})\hat H\hat\Psi_\alpha({\bf x)}\!+\!\mu_\alpha({\bf x})\hat\Psi_\alpha^\dagger({\bf x})\hat\Psi_\alpha({\bf x})}\\ \\
& & +\left[\Omega({\bf x})\hat\Psi_b^\dagger({\bf x})\hat\Psi_a({\bf x})+h.c.\right]
\Bigg\}.
\end{array}
\end{equation}
The ``chemical potentials'' $\mu_\alpha({\bf x})$ ($\alpha=a,\,b$) account for the two-photon detuning $2\delta$ and the Stark shifts arising from the two lasers $\Omega_1({\bf x})$ and $\Omega_2({\bf x})$. Hereafter we omit the spatial dependence of the chemical potentials as this will not change the conclusions, i.e. $\mu_\alpha({\bf x})\rightarrow\mu_\alpha$. The effective coupling between the two internal atomic levels is given by $\Omega({\bf x})=\frac{\Omega_1^*({\bf x})\Omega_2({\bf x})}{\Delta}$. Disorder in the present model derives from the spatial dependence of $\Omega({\bf x})$ which fluctuates both in phase and amplitude from site to site.

We proceed along the standard line~\cite{rev} and expand the field operators in in the single-band site-localized Wannier functions $w_{\alpha{\bf i}}({\bf x})$; $\hat\Psi_\alpha({\bf x})=\sum_{\bf i}\hat \alpha_{\bf i} w_{\alpha{\bf i}}({\bf x})$ with ${\bf i}$ the site index and $\hat a_{\bf i}$ and $\hat b_{\bf i}$ annihilate an $a$ and $b$ species atom at site ${\bf i} = (i_x, i_y)$ respectively. By further imposing the tight-binding approximation, the Hamiltonian becomes
\begin{equation}\label{ham1}
\begin{array}{lll}
\hat H_\mathrm{eff} & = & \displaystyle{-t\sum_{\langle{\bf ij}\rangle}\left(\hat a_{\bf i}^\dagger\hat a_{\bf j}+\hat b_{\bf i}^\dagger\hat b_{\bf j}\right)+\sum_{\bf i}\left(\mu_a\hat n_{a{\bf i}}+\mu_b\hat n_{b{\bf i}}\right)}\\ \\
& & +\displaystyle{\sum_{\bf i}h_{\bf i}\left(e^{i\varphi_{\bf i}}\hat a_{\bf i}^\dagger\hat b_{\bf i}+h.c.\right)}.
\end{array}
\end{equation}
Here the first sum includes only nearest neighbour terms, and we have neglected Raman-induced couplings between atoms in neighbouring sites. The atom number operators $\hat n_{a{\bf i}}=\hat a_{\bf i}^\dagger\hat a_{\bf i}$ and $\hat n_{b{\bf i}}=\hat b_{\bf i}^\dagger\hat b_{\bf i}$, and the onsite couplings are $h_{\bf i}=\left|\int d{\bf x}\,w_{a{\bf i}}^*({\bf x})w_{b{\bf i}}({\bf x})\Omega({\bf x})\right|$ and $\varphi_{\bf i}=\mathrm{angle}\left[\int d{\bf x}\,w_{a{\bf i}}^*({\bf x})w_{b{\bf i}}({\bf x})\Omega({\bf x})\right]$. These are random numbers taken from Gaussian distributions with widths 
$\zeta$ and $\xi$. Letting $\zeta=\xi=0$ we recover the clean system, and for $\xi=\infty$, which will be assumed throughout, the phases $\varphi_{\bf i}$ are uniformly distributed over $2\pi$. The phase diagram of the clean model has been studied in the past when atom-atom interaction was included~\cite{jonas1}, but the disordered coupled case is to the best of our knowledge so far unexplored even for zero interaction. 

The Hamiltonian (\ref{ham1}) will be the starting point for the analysis in Sec.~\ref{resultsec}. Before presenting the results, however, the various symmetry classes realizable with $\hat H_\mathrm{eff}$ are discussed next.

\subsection{Symmetries}\label{symmetrysec}
To date, the internal structure of the atoms plays no particular role in experiments of disorder with systems of cold atoms. Thus, the applied fields are very off-resonant from any atomic transitions and only the resulting Stark shifts generate the disordered potential. It is clear that the coupled system becomes much richer in the sense that a set of different models can be easily monitored by tuning the laser parameters. In particular, in such generalized situations the effective models carries an intrinsic pseudo-spin degree of freedom. The idea of this section is to present the various possible cases that can be achieved and to relate them to the corresponding symmetry class. A summary is given in the Tab.~\ref{tab:tab1} at the end of the section.

\subsubsection{Real-valued disorder and vanishing chemical potential; Class BDI}
The simplest disordered situations appear when the two chemical potentials are equal, i.e. we set $\mu_a=\mu_b=0$, and the disorder is purely real, $\varphi_{\bf i}=0$ ($h_{\bf i}$ can still change from site to site). By introducing the notation
$\hat A_{\bf i}=\left[\hat a_{\bf i},\,\,\hat b_{\bf i}\right]^t$ the Hamiltonian can be written as
\begin{equation}\label{h_BDI}
\hat H_\mathrm{BDI} = -t\sum_{\langle{\bf ij}\rangle}\hat A_{\bf i}^\dagger\hat A_{\bf j}
+\sum_{\bf i}h_{\bf i}\hat A_{\bf i}^\dagger\hat\sigma_x\hat A_{\bf i},
\end{equation}
where $\hat\sigma_x=\left[\begin{array}{cc} 0 & 1 \\ 1 & 0\end{array}\right]$ is the $x$-Pauli matrix. Now defining 
\begin{equation}\label{trans}
\hat c_{\bf i}=\frac{1}{2}\left(\hat{a}_{\bf i}+\hat b_{\bf i}\right),\hspace{1cm}\hat d_{\bf i}=\frac{1}{2}\left(\hat a_{\bf i}-\hat b_{\bf i}\right),
\end{equation}
and $\hat C_{\bf i}=\left[\hat c_{\bf i},\,\,\hat d_{\bf i}\right]^t$, Eq.~(\ref{h_BDI}) becomes
\begin{equation}\label{andmod}
\hat H_\mathrm{BDI} = -t\sum_{\langle{\bf ij}\rangle}\hat C_{\bf i}^\dagger\hat C_{\bf j}
+\sum_{\bf i}h_{\bf i}\hat C_{\bf i}^\dagger\hat\sigma_z\hat C_{\bf i}.
\end{equation}
In this transformed basis, this model is equivalent to the original Anderson Hamiltonian~\cite{anderson} in 2D, although here with two copies that experience the random potential with the same magnitude but opposite signs. This system has chiral symmetry~\cite{Chalker} and therefore the spectrum is symmetric around the zero energy. Specifically, in the representation of the $\hat A_{\bf i}$ operators introduced above, and noticing that the dimension is $2L^2$ where $L$ is, as before, the size of the lattice in one direction, we have for $L$ even that the operator~\cite{combase} 
\begin{equation}\label{anti}
\hat U=\mathrm{diag}(+1,-1,+1,\dots,-1|-1,+1,-1,\dots,+1)
\end{equation}
anti-commutes with the Hamiltonian. Moreover, since the Hamiltonian is real it is also time-reversal symmetric. Thus, the model belongs to the class BDI - of {\it chiral orthogonal} systems~\cite{haake,altland}.

\subsubsection{Complex-valued disorder and vanishing chemical potential; Class AIII}
Generalizing the previous case to complex couplings, $\varphi_{\bf i}\neq0$, the model is written
\begin{equation}
\hat H_\mathrm{AIII} = -t\sum_{\langle{\bf ij}\rangle}\hat A_{\bf i}^\dagger\hat A_{\bf j}
+\sum_{\bf i}h_{\bf i}\hat A_{\bf i}^\dagger\left(\cos\varphi_{\bf i}\hat\sigma_x+\sin\varphi_{\bf i}\hat\sigma_y\right)\hat A_{\bf i}.
\end{equation}   
Contrary to the case with real disorder, here we cannot decouple the two species with a spatially-independent transformation like Eq.~(\ref{trans}). Mixing in the $\hat\sigma_y$ component does not break the chiral symmetry, as can be seen from the fact that $\hat\sigma_z$ anti-commutes with $\hat\sigma_x$ and $\hat\sigma_y$, and the operator of Eq.~(\ref{anti}) has the form $\hat U=\hat D\otimes\hat\sigma_z$ with an $L^2\times L^2$ matrix $\hat D=\mathrm{diag}(+1,-1,+1,\dots,-1)$. The Hamiltonian is, however, complex and time-reversal symmetry is therefore broken. The symmetry class is AIII - of {\it chiral unitary} systems~\cite{haake,altland}.

An interesting observation is that by transforming the model with the unitary $\hat U_\varphi=\prod_{\bf i}\exp\left[-i\varphi_{\bf i}\left(\hat n_{a{\bf i}}-\hat n_{b{\bf i}}\right)/2\right]$, the random phase appears on the tunneling terms;
\begin{equation}\label{fluxm}
\hat H_\mathrm{AIII}'=\hat U_\varphi\hat H_\mathrm{AIII}\hat U_\varphi^{-1}= -\sum_{\langle{\bf ij}\rangle}\hat A_{\bf i}^\dagger\hat t_{\bf ij}\hat A_{\bf j}
+\sum_{\bf i}h_{\bf i}\hat A_{\bf i}^\dagger\hat\sigma_x\hat A_{\bf i},
\end{equation}
with
\begin{equation}
\hat t_{\bf ij}=t\left[
\begin{array}{cc}
e^{i\left(\varphi_{\bf i}-\varphi_{\bf j}\right)} & 0\\
0 & e^{-i\left(\varphi_{\bf i}-\varphi_{\bf j}\right)}
\end{array}\right].
\end{equation}
This is a version of the random flux model~\cite{rfm1,rfm} describing a random flux through any plaquette in the lattice. Thus, the spatially dependent phase $\varphi(x)$ of the Raman coupling effectively generates a synthetic magnetic field for the neutral atoms~\cite{gauge}.

\subsubsection{Real-valued disorder and no-zero chemical potential; Class AI}
Adding a chemical potential to the real-valued disorder case breaks the chiral symmetry, but the time-reversal one remains intact. The resulting Hamiltonian
\begin{equation}
\hat H_\mathrm{AI} = -t\sum_{\langle{\bf ij}\rangle}\hat A_{\bf i}^\dagger\hat A_{\bf j}
+\sum_{\bf i}\hat A_{\bf i}^\dagger\left(h_{\bf i}\hat\sigma_x+\mu\hat\sigma_z\right)\hat A_{\bf i},
\end{equation}
where we have introduced the effective chemical potential $\mu=(\mu_a-\mu_b)/2$ (we have subtracted a trivial term $(\mu_a+\mu_b)/2\sum_{\bf i}\hat A_{\bf i}^\dagger\hat A_{\bf i}$ from the Hamiltonian). With only time-reversal symmetry present, the model belongs to the class AI - of {\it Wigner-Dyson orthogonal} systems~\cite{haake,altland}.

\subsubsection{Complex-valued disorder and non-zero chemical potential; Class A}
By breaking both the chiral and the time-reversal symmetries, i.e., by including a chemical potential and considering a complex Raman coupling, respectively, the model belongs to the A class - of {\it Wigner-Dyson unitary} systems~\cite{haake,altland}. The Hamiltonian is then written
\begin{equation}
\hat H_\mathrm{A} \!=\! -t\!\sum_{\langle{\bf ij}\rangle}\!\hat A_{\bf i}^\dagger\hat A_{\bf j}
+\!\sum_{\bf i}h_{\bf i}\hat A_{\bf i}^\dagger\!\left(\cos\varphi_{\bf i}\hat\sigma_x+\sin\varphi_{\bf i}\hat\sigma_y+\mu\hat\sigma_z\right)\!\hat A_{\bf i}.
\end{equation}   

The underlying four symmetry classes of our general model Hamiltonian~(\ref{ham1}) are listed in Tab.~\ref{tab:tab1}.  
\begin{table}
\begin{tabular}{|c|c|c|}
\hline
 $h_{\bf i}$ &$\mu$ & Class\\
\hline
real-valued& zero & BDI (chiral orthogonal)\\
complex-valued& zero & AIII (chiral unitary)\\
real-valued& non-zero & AI (Wigner-Dyson orthogonal)\\
complex-valued & non-zero&A (Wigner-Dyson unitary)\\
\hline
\end{tabular}
\caption{Classification of Eq.~(\ref{ham1}) for different
  choices of $h_i$ and $\mu$ (see text for details).}
\label{tab:tab1}
\end{table}

\section{Results and discussions}\label{resultsec}
\subsection{Universal properties}\label{propsubsec}
Having identified the symmetry classes, it is now possible to directly determine some properties of the different models that can be realised from~(\ref{ham1}). The class AI is trivial in the sense that all states are localized and there is no room for 
non-trivial topological states, as e.g. delocalized edge states~\cite{2ddis}. In the symmetry class A, the states are also all localized, but the model can be potentially topological~\cite{2ddis}. The two other classes, the chiral ones, are more interesting in terms of the localization properties.
 
Chiral structures usually arise when a lattice system can be equivalently decomposed into two sublattices. The present model is clearly of this type, as the two species can be thought of as the constituents of a bilayer lattice, with a random tunneling $h_{\bf i}$ between the layers. In the usual classification scheme, there are three chiral classes: AIII, BDI and CII. As mentioned above, in AIII time-reversal symmetry is broken, while in BDI it is preserved and the time-reversal symmetry operator in this class squares to +1. The class CII is also time-reversal symmetric, but here the square of the time-reversal symmetry operator is -1, i.e. it is {\it chiral symplectic}~\cite{haake}. The symplectic classes typically contain a spin degree of freedom, and will not be of importance here. One can imagine, however, that by using four species (four different electronic Zeeman levels) this class could also be realized with our model.  

Although it has been shown that weak localization is absent in all orders in perturbation theory in chiral systems~\cite{weak}, recent studies support the possibility of a metal-insulator transition at the centre of the spectrum (for energies $E\sim 0$)~\cite{2dis2} as a function of the disorder strength. We thus expect a diverging localization length $\lambda$ for not too strong values of the disorder as the energy approaches zero in the chiral cases. In particular, for Gaussian disorder, a renormalization group analysis gives the localization length~\cite{2ddis}
\begin{equation}\label{loclength}
\lambda(E)\propto e^{g^{-1}\left|\log\left(E/\Delta\right)\right|^{1/2}},
\end{equation} 
where $g^{-1}$ is proportional to the conductance and $\Delta$ is the band width (of the clean system). It follows that for any finite energy $E$ the system is localized, albeit the localization length can get very long (in comparison to the system size here denoted as $L$).
According to the expression (\ref{loclength}), the system can be metallic in a strict sense only for $E\equiv0$, which implies that for finite systems the total number of sites has to be odd. Nevertheless, when the localization length $\lambda\gg L$, more states at the centre of the spectrum will always appear metallic in this case. In the infinite system, the analysis via the effective non-linear-$\sigma$-model~\cite{2ddis} from which Eq.~(\ref{loclength}) is derived, hints that indeed the $E = 0$ state should always be delocalized/metallic. However, Eq.~(\ref{loclength}) can also be obtained from a modified action, including higher-order terms, which shows that the $E=0$ state can be localized even for the chiral classes. This allows, therefore, for a metal-insulator transition. This transition is of the Kosterlitz-Thouless type and has a topological origin in terms of creation of vortex excitations which are responsible for the insulating phase~\cite{2dis2}.

Chirality is however broken as soon as $\mu\neq 0$, and the properties of the system are determined by the A or AI classes. This explains, at least qualitatively, why the states have to be localized for the non chiral cases considered in this work.

\subsection{Numerical results}\label{numsubsec}
As we have seen, we have at our disposal a system Hamiltonian (\ref{ham1}) where the parameters can be tuned such that four different symmetry classes can be explored. The great flexibility of the model compared to earlier studies of localization in cold atom settings is the multi-species structure. Thus, in the limit of large chemical potential $\mu=|\mu_a-\mu_b|/2$ (when one species is almost completely unpopulated) we recover the standard scenario of the Anderson model which has served as a theoretical model of most disordered cold atom experiments~\cite{atomloc,atomdis1,atomdis2,boson3d}. We should mention that even in the absence of interactions, as is the model studied here, the conclusions drawn from the numerical results should be taken with care;
the system is unavoidably finite and as soon as the localization length becomes comparable to the system size boundary effects play an important role. For this reason, the metal/insulator transition, conjectured to happen for $E=0$, has not been verified numerically, for example. 

\subsubsection{Localization properties}
A measure of localization is the inverse partition ratio (IPR)~\cite{2ddis}
\begin{equation}\label{ipr}
\mathrm{IPR}(E)=\sum_{{\bf i},{\bf j}}|\psi_E({\bf i},{\bf j})|^4,
\end{equation} 
where $\psi_E({\bf i},{\bf j})$ is an eigenstate with energy $E$. This is considered in the real space basis, and $\mathrm{IPR}=N^{-1}$  for a maximally delocalized state, with $N$ the dimension of the Hilbert space, and $\mathrm{IPR}=1$ for a maximally localized state. We conclude therefore that if $\mathrm{IPR}\sim N^{-1}$ the localization length exceeds the system size $L$ (which in our case is the number of sites in one of the directions), and we expect the IPR to drop as we approach the negative tail of the spectrum. As taken here, the IPR is the second power of the particle density, i.e., $\sum_{{\bf i},{\bf j}}\rho_E^q({\bf i},{\bf j})$ with $\rho_E({\bf i},{\bf j})=|\psi_E({\bf i},{\bf j})|^2$ and $q=2$. Other powers of $q$ might also be of relevance in connection with multifractality properties of the wavefunction which have been thoroughly discussed in terms of the Anderson metal-insulator transition~\cite{2ddis,multi}. In general, for given $q$ one finds a scaling $L^{\tau(q)}$
where $\tau(q)$ is a $q$- and dimensional dependent exponent, but such considerations have been left out of the present study. 
 
\begin{figure}[h]
\centerline{\includegraphics[width=8cm]{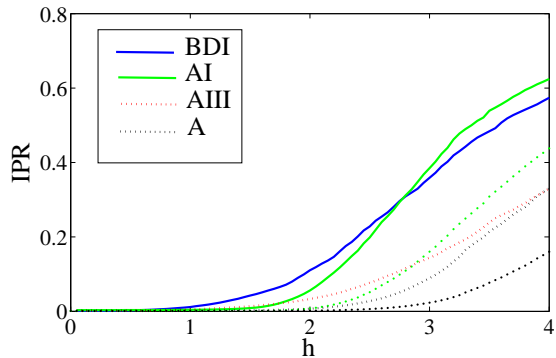}}
\caption{The inverse partition ratio IPR (\ref{ipr}) for the ground state of the Hamiltonian (\ref{ham1}) averaged over 100 disorder realizations in a $30\times30$ lattice. This size of the lattice is chosen from its relevance for experiments, and throughout this work we use periodic boundary conditions. The parameters are 
$t=1$, $\mu=(\mu_a-\mu_b)/2=2$, and $\xi=\infty$ and  the dotted lines display the IPR for the $b$ species atoms (clearly in the chiral classes the $a$ and $b$ species' IPRs are identical). As is seen, localization sets in at weaker disorder strengths for the chiral models in the ground state, and also for the low-lying excitations (not shown). As explained in the text, this result can be understood from an effective model for the non-chiral classes, which contains an inherent long-range (exponential) hopping.} \label{fig2}
\end{figure} 

Fig.~\ref{fig2} shows the IPR for the ground state of the four different symmetry classes as a function of the disorder strength $\zeta$. By comparing the models preserving or not preserving time-reversal symmetry, and chiral with non-chiral ones - that is, the BDI with AI, and the AIII with A, we notice that in the non-chiral classes A and AI, localization sets in later than in the chiral BDI and AIII in terms of the disorder strength. This result can be understood by considering Hamiltonian (\ref{ham1}) in the regime of $|\mu|\gg|t|,\,|h_{\bf i}|$ such that the two species can be effectively decoupled. In doing so, and integrating out over one of the species, the effective Hamiltonian describing the other one contains a tunneling term that is long range. 
The equations of motion for the atomic operators, given~(\ref{ham1}), obey
\begin{equation}\label{eom2}
\begin{array}{l} 
\displaystyle{\partial_t\hat a_{\bf i}=-ih_{\bf i}\hat b_{\bf i}-i\mu_a\hat a_{\bf i}+it\sum_{\bf j}\hat a_{\bf j}},\\ \\
\displaystyle{\partial_t\hat b_{\bf i}=-ih_{\bf i}\hat a_{\bf i}-i\mu_b\hat b_{\bf i}+it\sum_{\bf j}\hat b_{\bf j}}.
\end{array}
\end{equation}
Here, the sum $\sum_{\bf j}$ is over the nearest neighbours to the site ${\bf i}$. Let us assume that $\mu_a$ has the largest amplitude of all the parameters, and in this situations we set $\partial_t\hat a_{\bf i}=0$. In vector notation, the steady state solution for the species $a$ can be written $\hat {\bf a}={\bf M}^{-1}{\bf h}\hat{\bf b}$, with ${\bf h}$ a diagonal matrix with the diagonal elements $h_{\bf i}$ and ${\bf M}$ a tight-binding matrix with $-t$ as tunnelling strength and $\mu_a$ on the diagonal.
Inserting the steady state solution of $\hat a_{\bf i}$ into the second equation of (\ref{eom2}), it directly follows that the $a$ species atoms induce an effective long range hopping of the $b$ species ones. If $|\mu_a|<|t|$, the effective hopping is infinite range, while for $|\mu_a|>|t|$ it falls off as $\left(t/\mu_a\right)^{-|{\bf i}-{\bf j}|}$. Notice, however that this procedure is only consistent with the second case ($|\mu_a|>|t|$) since it relies on the adiabatic assumption $\partial_t\hat a_{\bf i}\equiv0$. An alternative derivation using the path integral formalism of the above result can be found in Ref.~\cite{thesis}. The effective long-range hopping counteracts localization, and since provided that $|\mu_a|>|t|$ the hopping is exponential in this model, we still expect localization to set in but for stronger disorder~\cite{longrange}. 

We have just shown that for non-chiral classes it is possible to think of the second species as generating an effective long range hopping. But what about the chiral classes? For the BDI class, direct inspection of Eq.~(\ref{andmod}) shows that there exists a simple (disorder-independent) localized basis where the model is indeed of short range character in terms of hopping. For the AIII class this is not as clear, however,
since there is no spatially independent transformation decoupling the species in this case, and neither a scale separation allowing for the derivation of an effective model. 
\begin{figure}[h]
\centerline{\includegraphics[width=8cm]{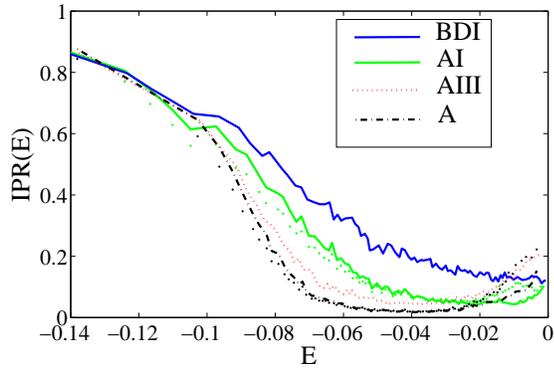}}
\caption{The inverse partition ratio $\mathrm{IPR}(E)$ (\ref{ipr}) as a function of the energy. Like in Fig.~\ref{fig2}, the IPR for the $b$ species atoms are marked by dotted lines. As argued in Subsec.~\ref{propsubsec}, the localization length grows as the energy $|E|$ is decreased. The flattening of the IPR is a result of the localization length $\lambda$ which becomes comparable or longer than the system size $L$ ($L=30$ for the lattice used here). The parameters are the same as those of Fig.~\ref{fig2} and with the disorder strength $h=10$, i.e. the low energy states are will localized.} \label{fig3}
\end{figure}

Let us now turn to localization of the excited states. As we discussed in the previous subsection, field theory methods have shown 
that the localization length $\lambda$ may diverge at $E=0$ in the chiral classes. For small but non-zero $|E|$, $\lambda$ is always finite, such that the states are localized - however possibly with a long localization length. A numerical check of Eq.~(\ref{loclength}) would require a system size $L$ far beyond of what is computationally achieved for our system. Indeed, as we see, even for energies not too close of $E = 0$ the localization length $\lambda\sim L$. When this happens $\mathrm{IPR}\sim N^{-1}$. The energy dependence of the IPR for the four classes is displayed in Fig.~\ref{fig3}. The ground and first excited states are clearly localized, but localization is quickly lost as the energy increases. The delocalized states appear more pronounced in the A and AIII models where time-reversal symmetry is broken. In these cases, the localization length becomes comparable to the system size already for the 100'th excited state (the total number of states is 1800). Closer to $E=0$, the IPR increases again, in contrast to what is analytically predicted in~(\ref{loclength}). We should point out, however, that Eq.~(\ref{loclength}) is obtained from an effective field theory that is valid only in the vicinity of $E=0$ and for infinite systems. As for the ground state, cf. Fig.~\ref{fig2}, localization is more pronounced in the models with chiral symmetry. It is tempting to argue that this is again a result deriving from the effective long-range hopping in these models. While this might be true, it should be noted that the population imbalance between the species is decreased for higher energies (still in the lower half of the spectrum) and for states at the center of the spectrum the two species are approximately equally populated.

\subsubsection{Properties of excited states}\label{sec:excitedstates}
Much less analytical results are available to describe the system away from the center of the spectrum. In systems of cold atom, with the high control of state preparation and of system parameters it is possible to study out-of-equilibrium dynamics following a quantum quench, see for example~\cite{quench}. By preparing an initial, non-stationary state with a given energy $\varepsilon>E_0$, where $E_0$ is the system ground state energy, and by properly adjusting the quench, the `energy window' scanned by the initial state could be made narrow~\cite{quench2}, and thereby properties of the excited states could be probed. 

The last term with the Raman coupling in Eq.~(\ref{ham1}) favours an onsite phase locking between the two species that is determined by the field phase $\varphi_{\bf i}$. Thus, if we are interested in the ground state and label it as 
\begin{equation}
|\psi_0\rangle=\left[
\begin{array}{c}
c_{a{\bf i}} \\
c_{b{\bf i}}\end{array}\right],
\end{equation}
where $c_{\alpha{\bf i}}$ ($\alpha=a,\,b$) is a vector indexed by the site subscript ${\bf i}$, then it follows that the energy of the coupling term is minimized whenever $\phi_{\bf i}\equiv\mathrm{angle}\left[c_{a{\bf i}}^*c_{b{\bf i}}\right]=-\varphi_{\bf i}+(2n+1)\pi$ for any integer $n$. 
In the classes with time-reversal symmetry, BDI and AI, $\varphi_{\bf i}$ is 0 or $\pi$ and the relative phase $\phi_{\bf i}$ is an integer multiple of $\pi$ at every site. Thus, the coefficients $c_{a{\bf i}}$ and $c_{b{\bf i}}$ will be real but may change sign (which also follows from the fact that the Hamiltonian is real and symmetric). Furthermore, in the chiral class BDI, $\phi_{\bf i}$ is independent on the site index ${\bf i}$ and according to Eq.~(\ref{andmod}), the eigenstates are found in either the $c$- or the $d$-subspaces. In the other two classes, AIII and A, the phase $\varphi_{\bf i}$ can be anything and consequently also the relative phase $\phi_{\bf i}$. Of course, the above picture is much simplified; whenever the phase $\phi_{\bf i}$ varies there is an additional cost of kinetic energy (this will be further discussed in the next subsection). Nevertheless, by combining the knowledge of the kinetic and coupling terms it is possible to determine qualitative properties of the excitations in the different models.

In the classes with time-reversal symmetry, the Hamiltonian is real and the eigenstates can always be chosen real. In other words, we can restrict the relative phase $\phi_{\bf i}$ to only integer multiples of $\pi$. The kinetic term is minimized if $\phi_{\bf i}=0$ or $\phi_{\bf i}=\pi$ for all ${\bf i}$'s. In the BDI class this means that the $\phi_{\bf i}$ has to be constant throughout the lattice, as a consequence of the decoupling of the species. This does not mean, however, that the phases of the individual species, i.e. of $c_{a{\bf i}}$'s and $c_{b{\bf i}}$'s, do not change between 0 and pi. Whenever this occurs it creates a domain wall in the wavefunction of the species in question. This implies an additional kinetic energy cost which is proportional to the length of the domain wall. Thus, we expect the domain wall to become longer for more excited states. This holds true only for not too highly excited states as we explain below. 

\begin{figure}[h]
\centerline{\includegraphics[width=4cm]{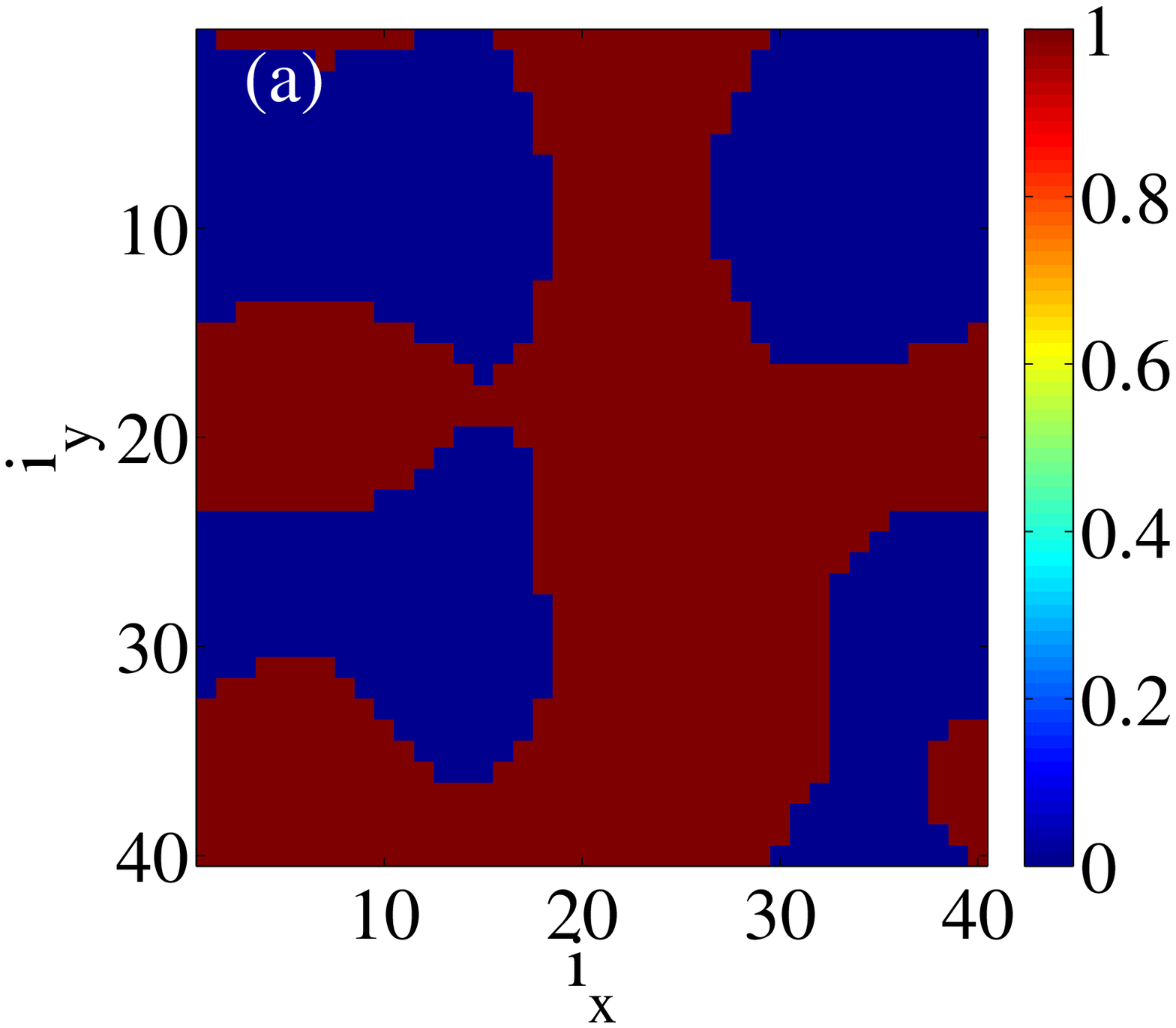} \includegraphics[width=4cm]{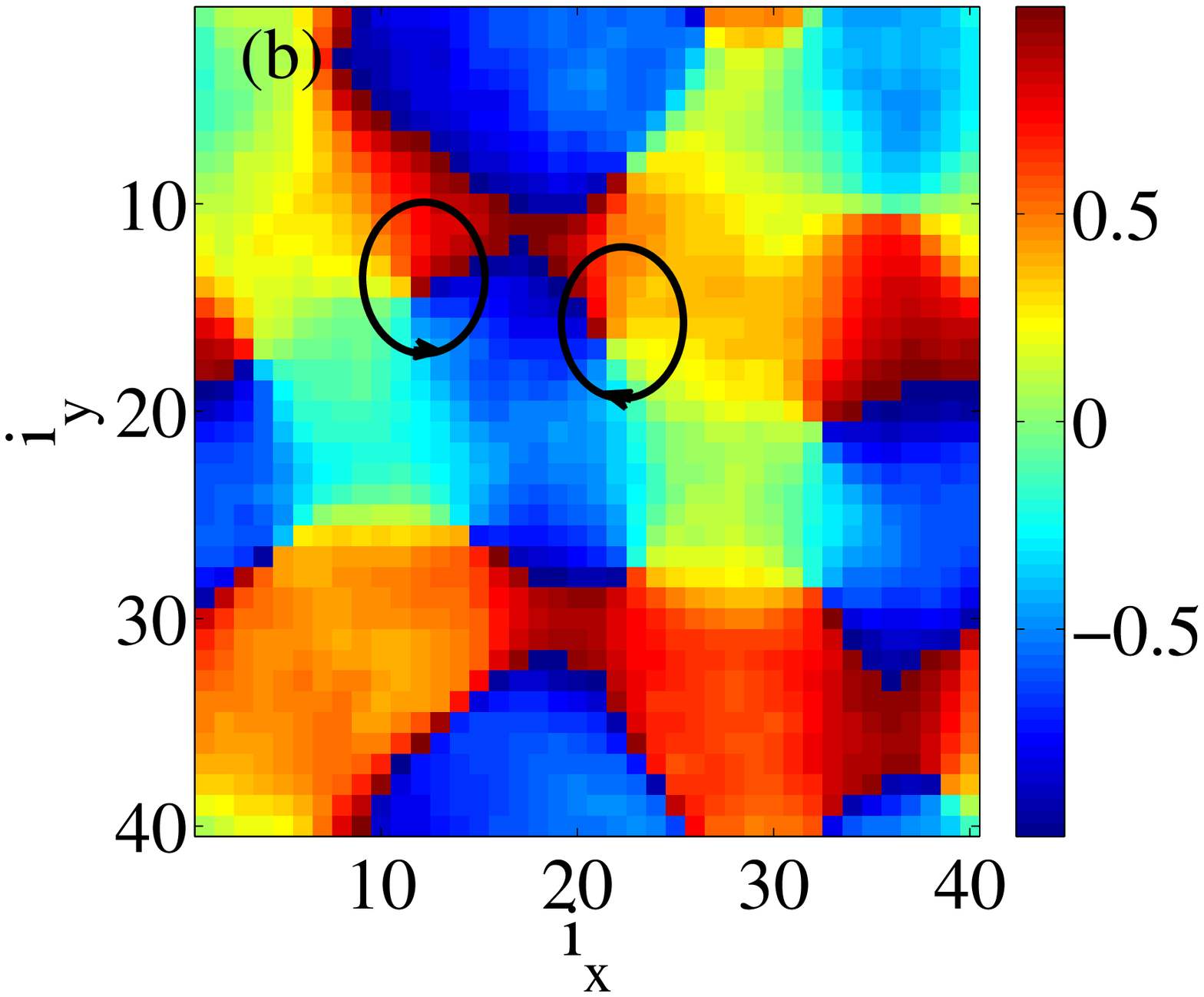}}
\caption{The individual phase for the $a$ species (i.e. the phase of the coefficients $c_{a{\bf i}}$) for the 20'th excited state for the class BDI (a) and AIII (b), and for a single disorder realization. In BDI the Hamiltonian is real and the only allowed excitations are domain walls, while in AIII the eigenstates are complex and vortices are typically formed. The parameters are 
$t=1$, $\mu=0$, $\zeta=0.4$, and $\xi=\infty$ in a $40\times40$ lattice.}\label{fig4}
\end{figure}

For the classes, AIII and A, the Hamiltonian is Hermitian and therefore its eigenstates are in general complex valued. Due to the kinetic energy cost, we expect the phase of the wave functions to vary smoothly between nearby sites. However, since the phase is defined modulo $2\pi$, the single valuedness of the wavefunction is still preserved for the $2\pi$ phase jumps, or in the presence of vortices (except at the vortex core were the phase is ill-defined and the density vanishes). The phase may thus contain branch cuts with a singularity at the ends of the cut, which reflects the vanishing density and the presence of vortices. In analogy with the cases with domain walls, we expect a larger number of vortices to appear higher up in the spectrum. In addition, since we consider finite lattices with periodic boundary conditions, any branch cut has two ends (unless it closes itself), and as a result a vortex is always accompanied with an anti-vortex. It might happen, though, that if the state is localized, then the vortex and the anti-vortex are within the region of non-zero particle density. As a result, experimentally it may happen that only one of the vortices in the pair is detected.

These conclusions are confirmed in Fig.~\ref{fig4} showing one example of the single species' phase for the BDI and the AIII classes. The first case displays a domain wall structure, while the appearance of vortices characterize the second case. Under periodic boundary conditions,  the domain walls are forced to close. We have indeed seen that the number of vortices and the length of the domain wall depends on the excitation energy as expected (we focus on energies $E<0$). We did not find, however, any particular relation that describes the increase in the number of vortices or domain walls as one goes up in the spectrum. Notice, in addition, that this increase in the number of vortices and in the length of the domain walls will only continue until roughly the centre of the spectrum. As mentioned above, this follows from the fact that in the chiral classes an eigenstate $\psi_E$ with energy $E$ has its companion state $\psi_{-E}=\hat U\psi_E$ with energy $-E$ and where $\hat U$ is the operator given in Eq. (\ref{anti}). This operator shifts the sign of the state on every second site. Therefore, if, say, the phase of $\psi_{-E}$ is smooth and positive (or negative), the phase of $\psi_E$ will have a checkerboard structure. In the AIII class we have also checked that all the vortices are characterized by winding number 1 and that no vortices with higher angular momentum are formed.

In experiments, we expect the vortices or domain walls to appear after a quench in the system. The vortices should be visible via time-of-flight and absorption detection techniques like in standard experiments~\cite{rot}. In order to single out vortices in only one of the atomic species, one can 
also envision state dependent detection. Likewise, the domain walls could be probed with the same methods; in a time-of-flight detection, a domain wall will for example be manifested as a zero-density cut in the momentum distribution.  

\begin{figure}[h]
\centerline{\includegraphics[width=4.4cm]{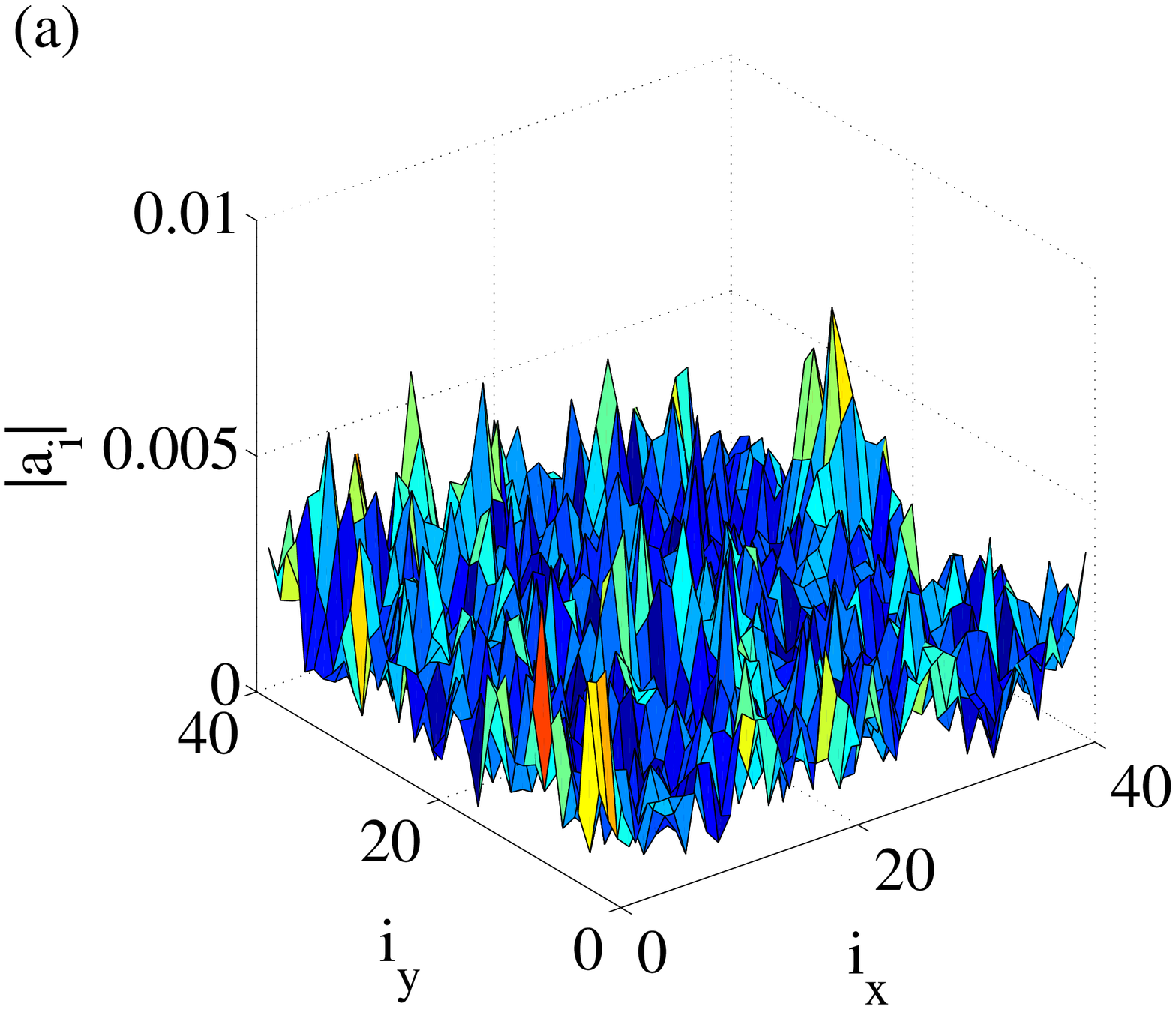} 
\includegraphics[width=4.4cm]{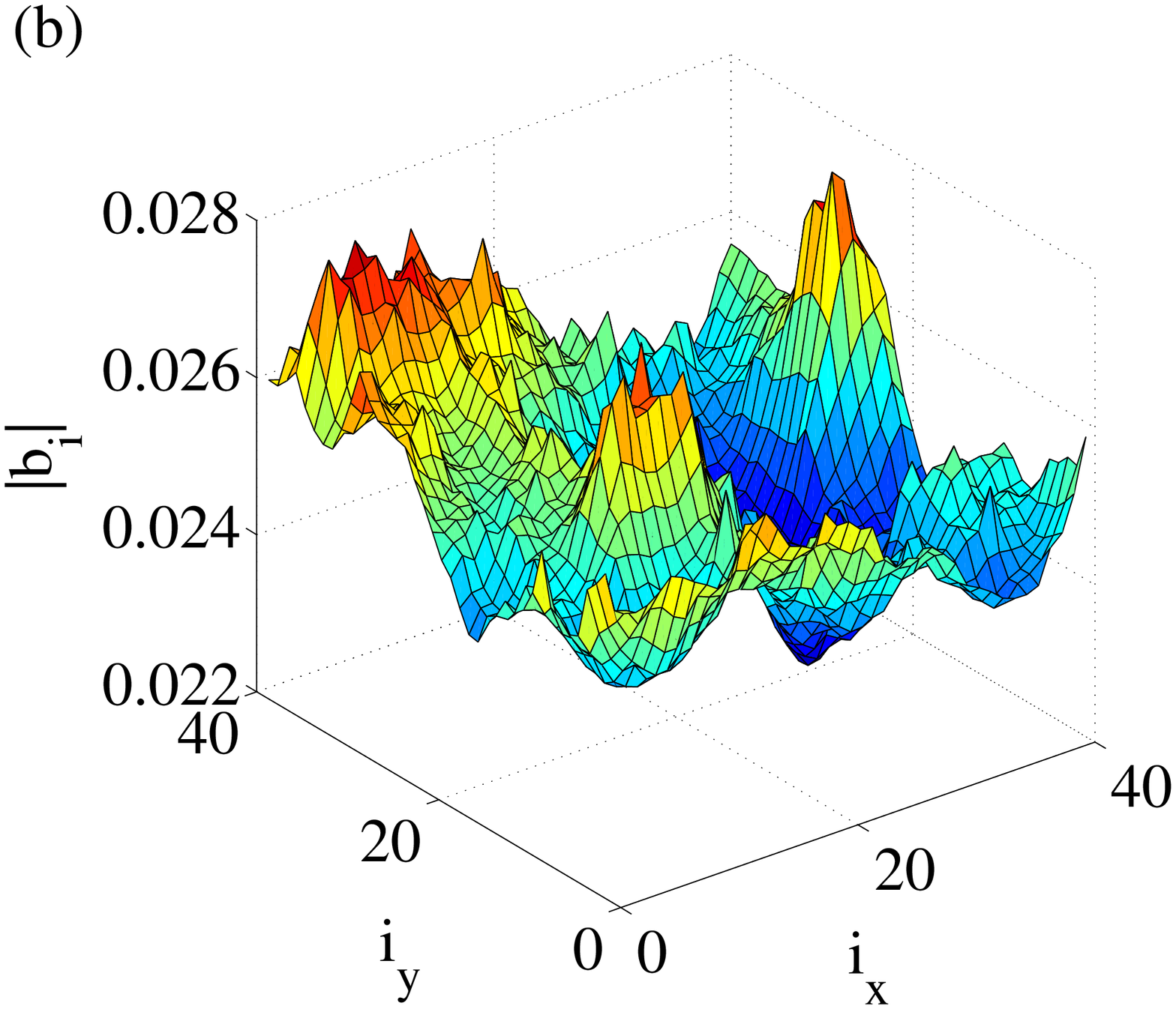}
}
\centerline{\includegraphics[width=4.2cm]{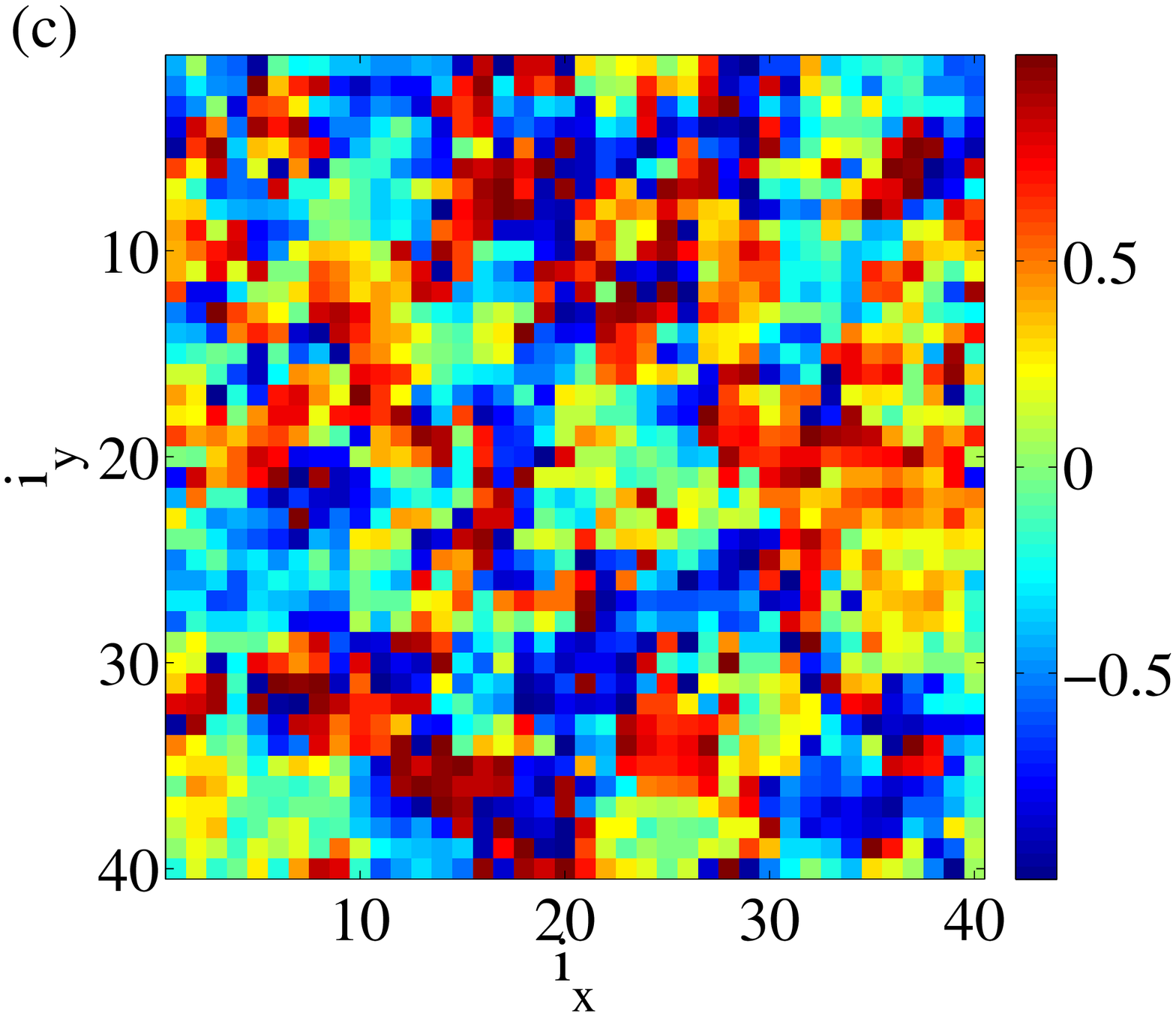}\includegraphics[width=4.2cm]{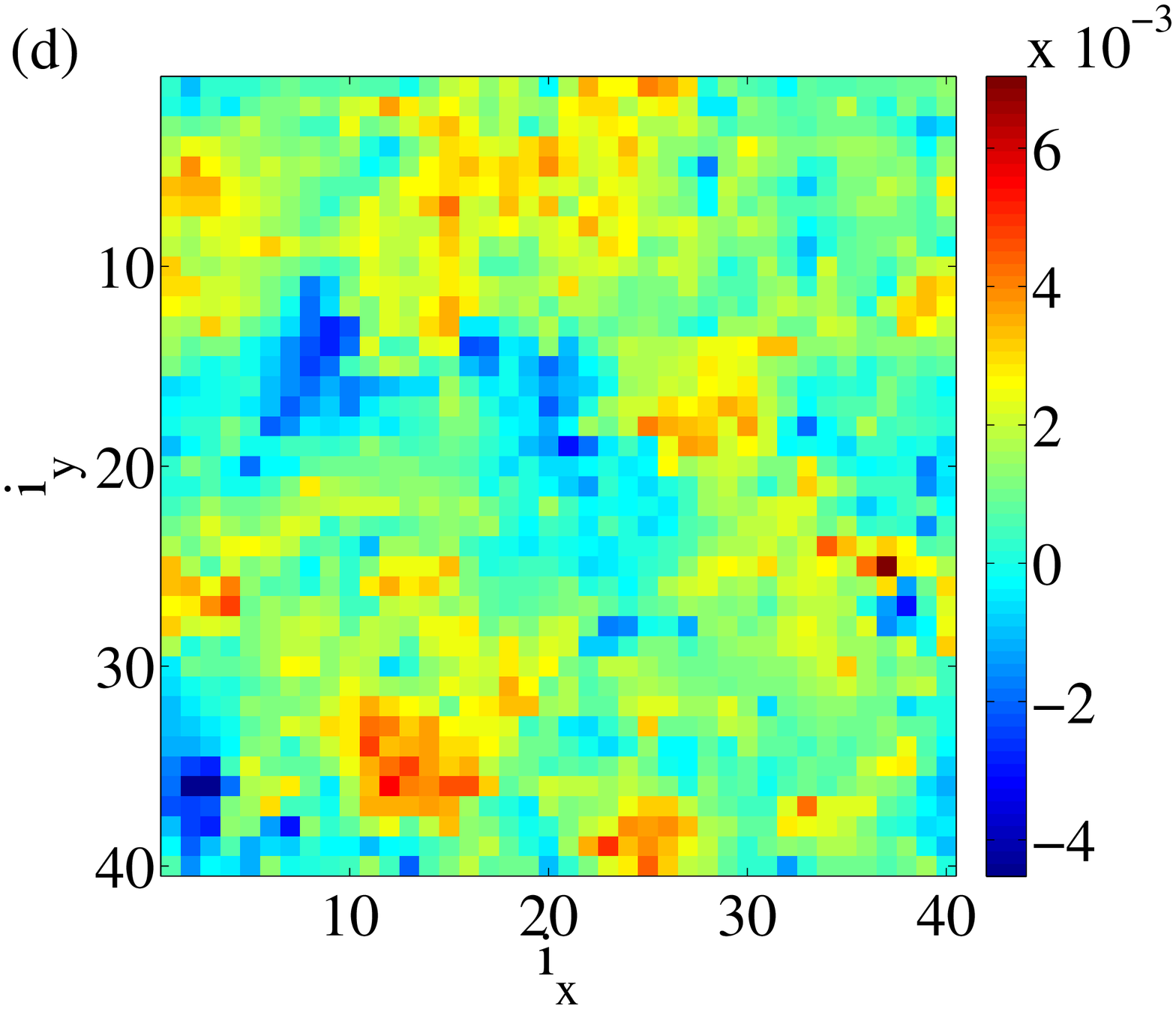}}
\caption{The densities (a) and (b), and the phases (c) and (d) for the ground state wave function in the non-chiral class A. As explained in the main text, the dominant species experiences a weaker effective disorder and its density thereby has a smooth envelope and its phase become constant, while the other species feels a strong effective disorder and its wave function then adjust accordingly. The parameters are the same as for Fig.~\ref{fig4} except $\mu=0.4$.} \label{fig5}
\end{figure}

For the above discussion of excited states we considered the chiral classes for which the $a$ and $b$ species are equally populated. For the non-chiral classes, A and AI, this equal balance is lost and for the low lying energy states the species with smallest chemical potential (we have defined the chemical potential as a positive energy cost) will dominate. As can be understood from the equations of motion (\ref{eom2}) this imbalance has a direct consequence on the structure of the states. If say $\mu_a>\mu_b\geq0$, and as long as the energy is not too big, mainly the $b$ species will be populated; $|\!|\hat a_{\bf i}|\!|<|\!|\hat b_{\bf i}|\!|$. Since the amount of disorder felt by one species is related to the population of the states of the other species, it follows that the atom of the $b$ type will experience a weak disorder, while the atoms populating the $a$-species states experience a strong disorder
(in comparison to the tunneling part). In Eq.~(\ref{eom2}) this is seen from the fact that the first terms on the right hand side are large in the upper equation and small in the lower equation. As a consequence, the wave function for the $b$ species will minimise the kinetic energy and its density will be smooth. The wave function for the $a$ species, on the other hand, will minimize the disordered part and the density and phase will show large fluctuations. This is displayed in Fig.~\ref{fig5} where we plot the ground-state densities and phases of both species.
Up in spectrum, the species with a higher chemical potential will get more populated, and excitations appear then partly as exciting the `internal' structure characterized by the two species. Finally, the situation is reversed in the highest excited states and population is predominantly occupying the $a$-species states. In addition, in the same way as for the chiral classes, the excitations in the classes where chiral symmetry is broken will also appear in the form of vortices and/or domain walls. 

\subsection{Random field induced order vs localization}\label{rfiosubsec}
The phenomenon known as the random-field induced order (RFIO) refers to the ability of certain systems to order only in the presence of a random field of a certain kind. 
This is the case when a clean system with a continuous symmetry in 2D, for example, is forbidden to order due to the restrictions of the Mermin-Wagner theorem~\cite{mw}, and then, after inclusion of a random field with discrete symmetry, the system does not fulfil the assumptions under which the theorem is valid, such that order is in principle not precluded. This has been proven to happen in the classical $X\!Y$ model~\cite{rfio}, and has also been suggested as the mechanism of order in graphene quantum Hall ferromagnets~\cite{Levitov}. 

In the RFIO, order appears in a particular way: In the ferromagnetic $X\!Y$ model in 2D, for example, the inclusion of a random field in the $x$-component of the spin will induce a ferromagnetic state with long-range order, but in the direction perpendicular to that of the field. In this case, the expectation value is finite for magnetization in the $y$ component. This can be understood from an intuitive argument, that the system minimizes the energetic cost of the random field by orienting all the spins in a direction in which the field exerts minimal influence - therefore the perpendicular choice~\cite{Levitov}. But despite the activity during the last years in the study of RFIO, the mechanisms allowing for such phenomenon are still not fully understood. A step in this direction, however, has been taken in~\cite{rfio}, where the occurrence of RFIO in the 2D $X\!Y$ model has been suggested in connection with the Anderson localization of spin waves.

This phenomenon has been also shown to happen in interacting systems of two-species Bose-Einstein condensates that are randomly coupled by a real-valued Raman field in~\cite{armand}. Here RFIO appears as a $\pi/2$ phase locking between the two species, and an experimental setup for observation has been proposed. Motivated by this, we have investigated Eq.~(\ref{ham1}) in the context of the RFIO. The advantage of the non-interacting system over the interacting one is the possibility of a direct check of localization properties in all the excited states. Therefore, in analogy to the system studied in~\cite{armand}, we expected RFIO to appear as a phase locking between the two species in the system of the class BDI, and following the reasoning, localization of the excitations should exhibit a different behavior from what was observed in the system of the AIII class, where the random Raman coupling only magnifies effects of the continuous symmetry.

As we discussed in subsection~\ref{sec:excitedstates}, all the states of the chiral BDI class have a homogeneous phase locking,
whereas this does not happens in the AIII class. This phase locking, however, does not seem to be a manifestation of the RFIO. First, because Eq.~(\ref{andmod}) is always real, and there is no reason for the appearance of a $\pi/2$ relative phase. Second, because since the relative phase is always locked (at random) at $0$ or $\pi$, a $\pi/2$ relative phase would be only the result of averaging, and therefore would not be relevant for experiments with single disorder realizations. In addition, since all low-lying states are localized in both the BDI and the AIII classes, the existence of phase coherence in the BDI one does not seem to have any relation with the localization of the excitations. 
This means that even if the homogeneous phase could be a manifestation of the RFIO, which appears in a particular and still unidentified way for the system discussed here (see Eq. (5) in~\cite{armand}), it would nevertheless be unrelated to the localization of the low lying excited states in this case. 

The close relation between the present lattice model and the continuum model of Ref.~\cite{armand}, and the fact that we do not find the expected signatures of RFIO, suggest that interaction may play a crucial role behind RFIO. This would have direct consequences on the hypothesis that understanding better the localization of excitations could also explain the origin of RFIO. In fact, the mechanisms behind localization in single-particle or interacting many-particle systems are very different. Anderson localization can be understood from interfering loops in configuration space, and especially the loops may be very long. In an interacting system, the destructive interference loops occur instead in a hyper cubic lattice and the loops are typically extremely short~\cite{altshuler}. Our results illustrate once again~\cite{rfio} that the mechanisms allowing for the phenomenon of RFIO constitute a very interesting open question.

\section{Conclusion}\label{sec:con} 
In this work we studied a model consisting of two species in a square optical lattice. The two species represent internal atomic Zeeman levels that are Raman coupled with an amplitude $h_{\bf i}$ which randomly varies between the sites. The disorder resulting from the Raman coupling tend to localize the eigenstates of the system. However, the properties of the states depend strongly on the symmetry class of the model. In particular we showed that with this simple model it is possible to realize four different symmetry classes. In terms of the localization, two of them, the chiral ones, are especially interesting since they may support
 a transition to a metallic phase in the centre of the spectrum. Strictly speaking,
the metallic state should only appear at zero energy, but since the system always has a finite size, a metallic-looking phase should also appear at nonzero energies.  
Experimentally, such extended states could be studied after a quench were an initially localized state is prepared with an energy close to zero; if the system is metallic, then one should see an everlasting diffusion of the density, while in the Anderson insulating phase the diffusion is rapidly hindered following a short expansion.

Properties of the symmetry classes are also reflected in the sort of excitations. We found that excitations can be characterized as either domain walls or vortices depending on whether the Hamiltonian is hermitian or real and symmetric. It should be noticed that these results only follow due to the two-species structure of the problem, and that similar excitations cannot occur in similar models with only one species. We also argued that both the vortices and the domain walls should be verifiable in time of flight measurements following after a quench of the system, and we concluded with a discussion on the phenomenon of the RFIO.

The present model could easily be extended to other situations: for example, to include more species (internal electronic levels), different lattice geometries and dimensions, and to include atom-atom interaction. These ideas could also be explored to engineer systems in the CII class by using four-level atoms, were the levels are coupled such as to yield two species, $a$ and $b$, both with an intrinsic spin-1/2 structure. Probably most interesting, however, would be to consider the present model with interaction included. How to characterize interacting systems in  `periodic table' is a most open question, and here experiments might provide valuable information in order to understand the new physics.

\appendix


\begin{acknowledgements}
Dmitry Bagrets and Alexander Altland are thanked for helpful discussions about the chiral symmetry classes. Maciej Lewenstein is thanked for bringing up the idea about RFIO in the present model. The authors acknowledge financial support from VR-Vetenskapsr\aa set (The Swedish Research Council). JL acknowledges KAW (The Knut and Alice Wallenberg foundation), and FP thanks the hospitality of ICFO, where part of this research was developed.
\end{acknowledgements}


\begin{thebibliography}{999}
\bibitem{com1} We go to a rotating frame with respect to the frequency of the coupling lasers and neglect rapidly oscillating terms.
\bibitem{com2} Note that this effective Hamiltonian can alternatively be derived from a path integral approach where we integrate out the $c$-field and neglect any retardation effects motivated by the dispersive coupling regime, see Ref.~\cite{thesis}.
\bibitem{combase} In the expression (\ref{anti}) for the unitary operator $\hat U$ which anti-commutes with the Hamiltonian we have assumed an even number of sites in both directions. Furthermore, it is clear that the explicit form of $\hat U$ given in the equation is basis dependent.
\end{thebibliography}
\end{document}